\documentclass[pra,twocolumn]{revtex4}
\usepackage{bbm}
\usepackage{natbib}
\usepackage{graphics}
\usepackage{amsmath}
\usepackage{mathrsfs}
\usepackage{theorem}
\usepackage{float}
\usepackage{hyperref}
\usepackage{rotating}
\usepackage{amssymb}
\usepackage[english]{babel}
\usepackage{color}
\usepackage{fancybox}

\newcommand{\ket}[1]{|#1\rangle}
\newcommand{\bra}[1]{\langle #1|}
\newcommand{\inp}[2]{\langle #1 | #2\rangle}
\newcommand{\be}{\begin{eqnarray}}
\newcommand{\ee}{\end{eqnarray}}

\begin{document}

\title{Effective Hamiltonian approach to the quantum phase transitions in the extended Jaynes-Cummings model}
\author{H. T. Cui $^{1}$}
\email{cuiht01335@aliyun.com}
\author{Y. A. Yan $^{1}$}
\email{yunan@ldu.edu.cn}
\author{M. Qin $^{1}$}
\email{qinming@ldu.edu.cn}
\author{X. X. Yi $^{2}$}
\email{yixx@nenu.edu.cn}
\affiliation{$^1$ School of Physics and Optoelectronic Engineering \& Institute of Theoretical Physics, Ludong University, Yantai 264025, China}
\affiliation{$^2$ Center for  Quantum Sciences, Northeast Normal University, Changchun 130024, China}
\date{\today}

\begin{abstract}
The study of phase transitions in dissipative quantum systems based on the Liouvillian is often hindered by the difficulty of constructing a time-local master equation when the system-environment coupling is strong. To address this issue, the complex discretization approximation for the environment is proposed to study the quantum phase transition in the extended Jaynes-Cumming model with an infinite number of boson modes. This approach yields a non-Hermitian effective Hamiltonian that can be used to simulate the dynamics of the spin. It is found that the ground state of this effective Hamiltonian determines the spin dynamics in the single-excitation subspace. Depending on the opening of the energy gap and the maximum population of excitations on the spin degree of freedom, three distinct phases can be identified: fast decaying, localized, and stretched dynamics of the spin. This approach can be extended to multiple excitations, and similar dynamics were found in the double-excitation subspace, indicating the robustness of the single-excitation phase.
\end{abstract}


\maketitle

\section{introduction}
The study of open quantum systems has revealed that phase transitions can occur, even though the system has a finite number of degrees of freedom \cite{hwang2016}. Unlike the phase transitions in closed systems under thermodynamic conditions, the phase transitions in open quantum systems are accompanied by significant changes in the system's dissipative dynamics. Thus, the so called dissipative phase transition (DPT) reflects  the non-equilibrium of a system and  the nontrivial properties of the steady state of the system. Moreover, by environment engineering, one can prepare the system in a special state through DPT, that would be difficult or impossible by conventional ways \cite{diehl2008}. To characterize DPT explicitly, the system's Liouvillian eigenvalues are studied ,  of which the real part corresponds to the decay rate of eigenmodes \cite{kessler12}. It was noted that the vanishing of the gap in the excitation spectrum of Liouvillian indicates a non-analytical change in the dissipative dynamics of system, which typically signals the onset of a phase transition \cite{kessler12}.

Recent experimental advances have allowed physicists to investigate regimes  beyond Markovian approximation, such as in solid-state \cite{solid-state} and artificial light-matter systems \cite{light-matter}. In this situation, the Liouvillian is difficult to determine due to the  memory effect of the environment. This raise the question of whether  phase transitions can still be found in the absence of a Liouvillian and if so, how to characterize it explicitly. To address this,  the method of perturbational expansion or numerical simulation  has been used to incorporate the influence of the environment \cite{vega2017}. However, the resulting dynamical equations  are often very complicated, making it difficult  to determine the phase transition \cite{breuer2002}. An alternative approach  is to approximate the environment as a many-body system with  discrete and finite energy modes.  Thus, an effective Hamiltonian can be constructed to predicate the dynamics of system \cite{kazansky97, discretization}. However, this approach requires a large Hilbert space dimension  to accurately  predicate the long-term behavior of system, which would become unaffordable in computation. To surmount this difficulty,  a complex discretization approximation for the environment was proposed  by the authors \cite{cui23}, which allows for a more  efficient computation and more faithful depiction of the dissipative dynamics of system by introducing  an effective Hamiltonian.  Thanking to the non-Hermitianity  of effective Hamiltonian, the decay rate of the eigenmode can be characterized by the imaginary part of the corresponding eigenvalue.

The similarity of the Liouvillian and the effective Hamiltonian encourages us to explore the phase transition by solving the complex Hamiltonian. This approach incorporates  the dynamical information for both the system and environment. So by studying the  eigenvalues and corresponding eigenvectors,  one can gain deep insight into the dynamics of a system. As an illustration, the extended Jaynes-Cumming model is discussed in this paper. This model is  the  generalization of the spin-boson model under rotating-wave approximation. It was known that the spin-boson model undergoes a quantum phase transition in the sub-Ohmic or Ohmic regime from a delocalized phase with a single ground state and no magnetization, to a localized phase with two-fold degenerate ground state and a nonzero magnetization \cite{leggett}. However, the nonpreservation of total excitations in the spin-boson model makes the investigation of  the phase transition complicated \cite{sbmodel}. The rotating-wave approximation is a reasonable choice to obtain tractable solutions \cite{breuer2002}. Furthermore, it allows us to determine the phase transition  by finding the singularity of order parameter. Similar research was performed in Ref. \cite{wang19},  where the effective  Hamiltonian was constructed in real space. Thus, the magnetization $\langle \sigma_z \rangle$ and $\langle \sigma_x \rangle$ are evaluated for the ground state  to detect phase transitions.

In contrast, a non-Hermtian effective Hamiltonian is constructed in this paper to determine the phase transition in the single-excitation subspace of the extended Jaynes-Cumming model. Because of the non-Hermitianity, the eigenvalue of the effective Hamiltonian shows the imaginary part, which characterizes the decaying dynamics of spin. Importantly,   besides  the known decaying phase and localization phase \cite{kofman1994},  the intermediate phase can be verified through checking the ground state function.  The intermediate phase demonstrates stretched-like behavior in dynamics.  The approach can be readily extended into the multi-excitation subspace. As a result, we explore the spin dynamics in the double-excitation subspace by the exact diagonalizatin method. Interestingly, the spin dynamics in this case could show similar behavior as that in the single-excitation subspace, which implies that the single-excitation phases would be stable.

The remainder of the paper is organized as follows. The model and method are introduced in Sec. II, with particular  remarks on the complex discretization approximation for environment. In Sec. III, the spin dynamics  is discussed explicitly in the single-excitation subspace by solving the effective complex Hamiltonian. Resultantly, a phase diagram is founded.  In Sec. IV, we extend the discussion into the double-excitation subspace. It is found  that the spin dynamics in this case is nearly the same as in single-excitation subspace.  Finally, conclusions  are offered in Sec. V.

\section{Model and method}

In this section, we introduce the model and present the overview of the complex discretization approximation.

\subsection{Extended Jaynes-Cumming model with infinite number of boson modes}

Its Hamiltonian can be written as $(\hbar=1)$
\be\label{H}
H=\Delta \sigma_+ \sigma_- +\sum_k \omega_k a^{\dagger}_k a_k +\sum_k  \left( g_k a^{\dagger}_k \sigma_- + g^*_k a_k \sigma_+\right).
\ee
where $\sigma_{\pm}= \left(\sigma_x \pm \mathbbm{i} \sigma_y\right)/2$ with $\sigma_i (i=x, y, z)$ the Pauli matrices and $\mathbbm{i}$ is the imaginary unit. $ a^{\dagger}_k ( a_k)$ denotes the creation (annihilation) operator of mode $k$ in the environment, which is coupled to the spin via coupling strength $g_k$. In this paper, the spectral density for the environment is chosen as
\be\label{J}
J(\omega)= \sum_k \left| g_k\right|^2\delta\left(\omega-\omega_k\right)=\eta \omega \left(\frac{\omega}{\omega_c}\right)^{s-1}e^{-\omega/\omega_c},
\ee
where $\eta$ depicts the coupling strength, and $\omega_c$ is the cut-off of frequency. Depending on the value of $s$, the environment is classified as sub-Ohmic $\left( 0< s < 1 \right)$, Ohmic $\left( s=1 \right)$ or super-Ohmic  $\left( s>1 \right)$. Despite the simplicity of Eq. \eqref{H}, the exact spin dynamics can be obtained only in the single-excitation subspace. In this case, the state of the total system can be written as
\be \label{psit}
\ket{\psi(t)}=\alpha(t)\ket{e} \ket{0}^{\otimes N} + \ket{g} \left(\sum_{k} \beta_k(t) \ket{1}_k\right),
\ee
in which $N$ denotes the number of modes in the environment,  $\ket{1}_k = a^{\dagger}_k \ket{0}_k$ with the vacuum state $\ket{0}_k$, and $\ket{e}= \sigma_+ \ket{g}$.  Substituting Eq. \eqref{psit} into the Schr\"{o}dinger equation, one gets
\be\label{alphat}
\mathbbm{i}\frac{\partial }{\partial t}\alpha (t)&=& \Delta \alpha(t)- \mathbbm{i} \int_0^t \text{d}\tau \alpha(\tau)\int_0^{\infty} \text{d} \omega J(\omega)e^{-\mathbbm{i} \omega(t-\tau)},
\ee
where the last term stems from the memory effect of the environment.

It is known that the spin dynamics in the single-excitation subspace  can display the transition from decaying to localization because of the occurrence of a  bound state \cite{kofman1994, bs}. The  bound state corresponds to a singular energy level, separated  from the continuum of the  environment. The existence of bound staten can be verified by solving Eq. \eqref{alphat} \cite{kofman1994}.  By using Laplace transformation, Eq. \eqref{alphat} can be rewritten as
\be\label{alphap}
\alpha(p)= \left[\mathbbm{i}\Delta +p +\mathbbm{i}\int_0^\infty\text{d}\omega \frac{J(\omega)}{\mathbbm{i}p -\omega}\right]^{-1},
\ee
where $\alpha(p)=\int_{0}^{\infty}e^{-pt} \alpha(t)$. If $p=-\mathbbm{i}E $ satisfying the equation
\be\label{bs}
\Delta- E + \int_0^\infty\text{d}\omega \frac{J(\omega)}{E -\omega}=0,
\ee
one then obtains $\alpha(t) \propto  e^{-\mathbbm{i}Et}$ by inverse Laplace transformation. This feature means that the modulus of $\alpha(t)$ may be constant, leading to localization in the dynamics. The real solution to Eq. \eqref{bs} can be found in the regime of  $E<0$, depending on $J(\omega)$ \cite{bs}. Moreover, since $\omega_k\in \left[0, \infty \right)$, the bound state can be viewed as the ground state of the total system. In this sense, the transition is a result of the opening of energy gap of the ground state from the continuum.

Equation \eqref{alphat}  can be solved by using the iterative method, but the memory effect renders the calculation  time-consuming and exhaustive for long term evolution. For multiple excitations,  the lack of a dynamical equation similar to Eq. \eqref{alphat} renders it difficult to analyze the spin dynamics and determine the bound state \cite{shi16}.  To address this, a feasible approach is to discretize the environment into a finite set of modes, thereby constructing an effective Hamiltonian that can be solved with exact diagonalization \cite{discretization}. However, this approach suffers from the  recurrence caused by the finite dimension, which make the evaluation unaffordable for long term evolution.

\begin{figure}[tb]
\center
\includegraphics[width=5cm]{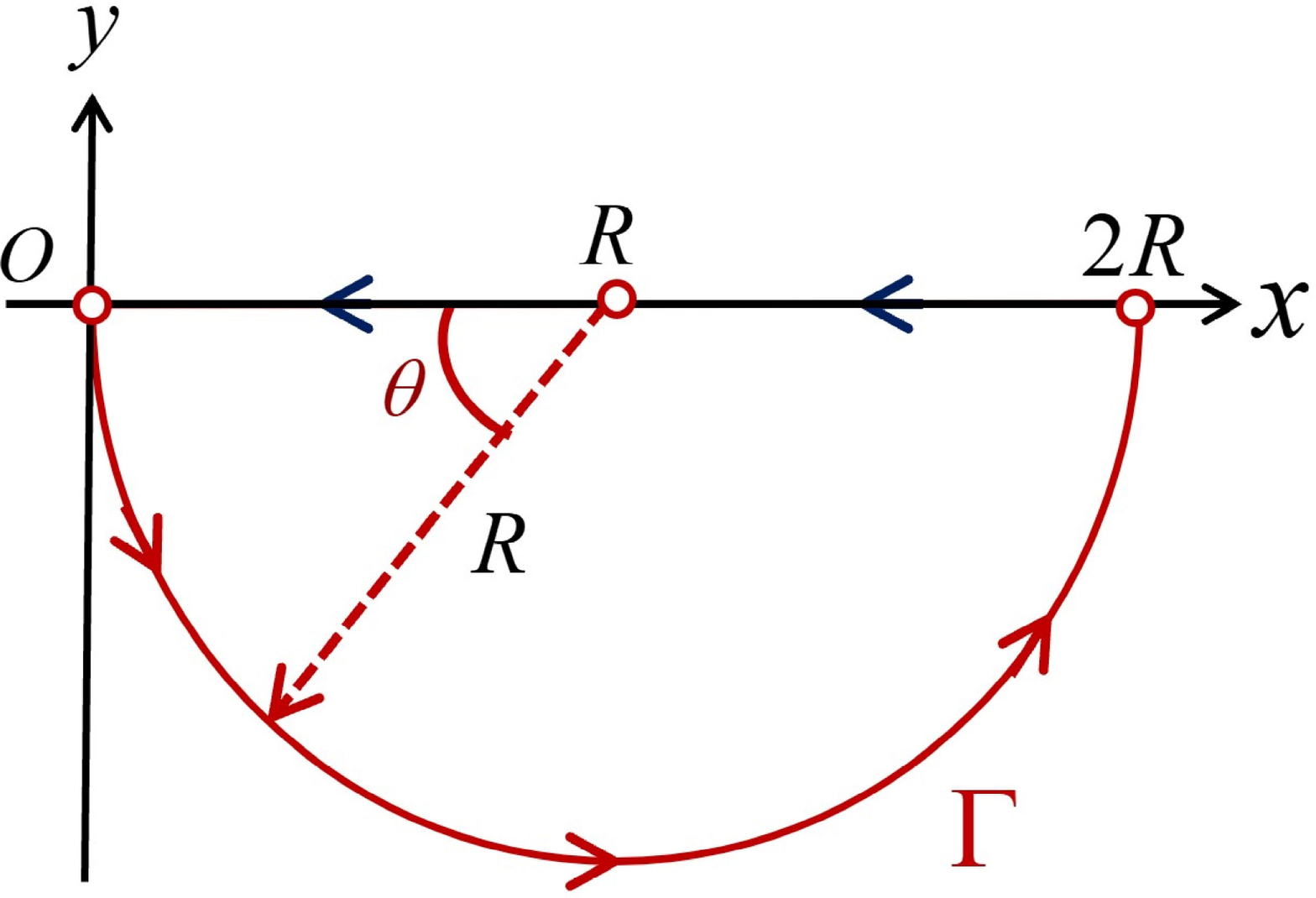}
\caption{(Color online)Illustration of the path in complex plane used to approximate  $\int_{0}^{\infty} \text{d}x$ as $\int_{\Gamma} \text{d}z$, where $\Gamma$ denotes the semi-circle with radius $R$, centered at $\left(R, 0\right)$. }
\label{fig:Gamma}
\end{figure}

\subsection{Complex discretization approximation}\label{section:CDA}

\begin{figure*}[tb]
\center
\includegraphics[width=16cm]{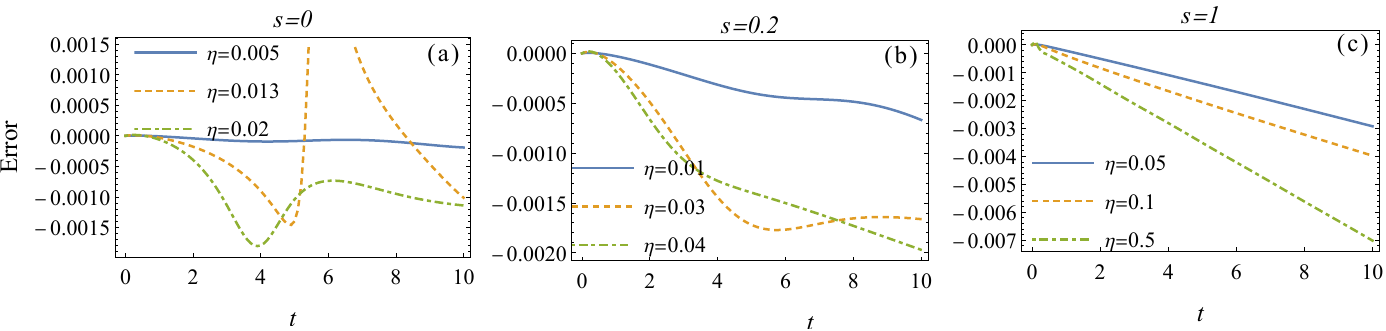}
\caption{(Color online)The plots for the computational errors defined in Eq. \eqref{error} vs. the evolution time $t$ (in units of $1/\Delta$).  For all plots, $N=2000, R=6, \Delta=1$, and $ \omega_c =10$ are chosen.  }
\label{fig:error}
\end{figure*}

\begin{figure}[tb]
\center
\includegraphics[width=6cm]{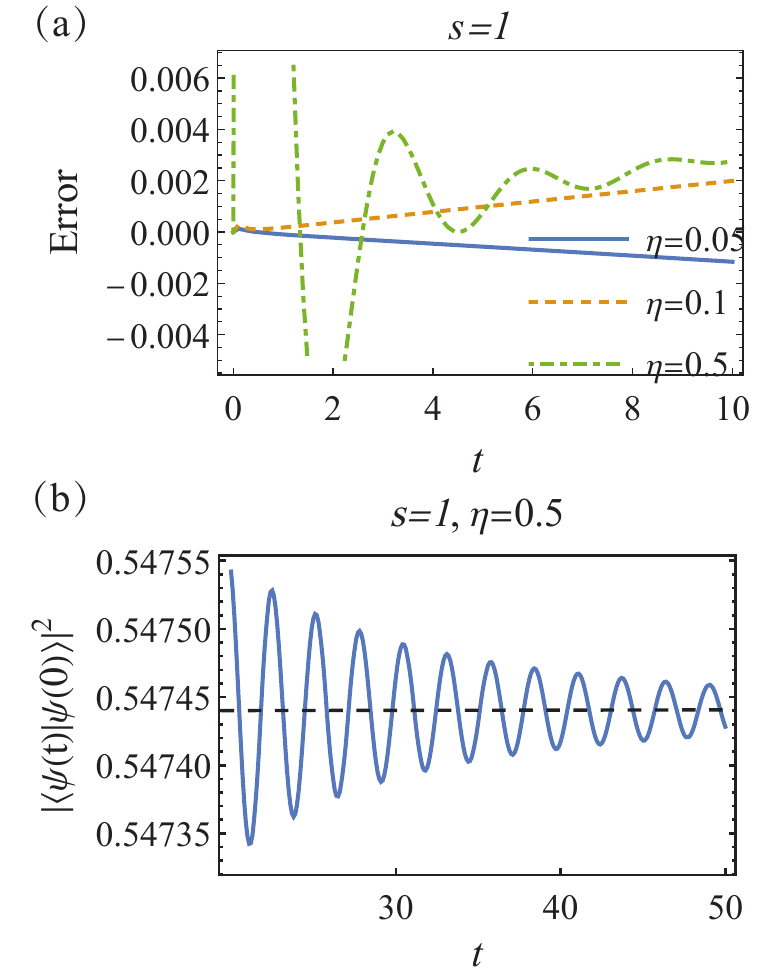}
\caption{(Color online)(a) The plot for the computational errors  vs. the evolution time $t$ (in units of $1/\Delta$) when $\widetilde{H}_{\text{dis}}$ is used to evaluate the survival probability for $s=1$. It is stressed that the data  for $\eta=0.05$ and $0.1$ are enlarged by ten times for convenience of illustration. The other parameters are the same as those in Fig. \ref{fig:error}. (b) The detailed plot for the oscillation of  survival probability when $\eta=0.5$. The dashed line denotes the value $0.54744$. The evaluation is obtained by $\widetilde{H}_{\text{dis}}$ with $N=2000$. The other parameters are the same as those in Fig. \ref{fig:error}. }
\label{fig:errorHtilde}
\end{figure}

Recently, the complex  discretization approximation (CDA) method was introduced to refine the simulation of spin dynamics \cite{cui23}. In contrast to its counterpart in real space \cite{discretization}, this method yields  a non-Hermitian effective Hamiltonian $H_{\text{dis}}$, which has complex eigenvalues with a negative imaginary part. Consequently, $H_{\text{dis}}$ can provide a more accurate description of the dissipative dynamics of spin.

The discretization approximation was proposed to tackle the infinite energy modes in an environment by utilizing  the Gauss quadrature \cite{discretization}. For this purpose,  Eq. \eqref{H} is transformed  under the continuous limit as \cite{bulla1997},
\be \label{continuousH}
H&=&\Delta \sigma_+ \sigma_- +\int_{0}^{\infty} \text{d}x h(x) a^{\dagger}_x a_x +\nonumber \\
&&\int_{0}^{\infty} \text{d}x   \left[g(x) a^{\dagger}_x \sigma_- +g^*(x)  a_x \sigma_+\right].
\ee
The operator $a_x$ or $a^{\dagger}_x$ fulfill the bosonic commutation relation $\left[a^{\dagger}_x, a_{x'}\right]=\delta(x -x')$.  The spectral function $J(\omega)$ can be rewritten for variable $x$ as \cite{bulla1997}
\be\label{sd}
J(x)= g^2\left[h^{-1}(x)\right]\frac{\text{d}h^{-1}(x)}{\text{d}x}.
\ee
The equivalence between Eqs. \eqref{H} and  \eqref{continuousH} is guaranteed by the fact that the same effective action for spin degree of freedom can be derived from the Hamiltonians Eqs. \eqref{H} and   \eqref{continuousH} \cite{bulla1997}.

It is noteworthy that the forms of  $h(x)$ or $g(x)$ are not uniquely determined for a given $J(x)$. By this freedom, one can choose proper form for  $h(x)$ or $g(x)$ for the sake of  discretization of the environment \cite{discretization}. Given the Ohmic spectral function Eq. \eqref{J}, one may choose
\be\label{gh}
h(x)=\omega_c x,
g(x)= \sqrt{\eta} \omega_c x^{s/2} e^{- x/2}.
\ee
Assuming that $h(x)$ has a linear relationship with variable $x$ is beneficial to the discretization approximation for  environment.

The main idea of CDA is to utilize complex Gauss quadratures to discretize the continuum of the environment. For this purpose, $\int_{0}^{\infty} \text{d}x$ is replaced by a line integral $\int_{\Gamma}\text{d}z$ in the complex plane shown in Fig. \ref{fig:Gamma}, where $z=x+\mathbbm{i}y$. $\Gamma$ denotes the semicircle with radius $R$ in the lower half of the complex plane, centered at $\left(R, 0\right)$. With the requirement that both $h(x)$ and $g(x)$ have no singularity in the complex plane, $\int_{\Gamma} \text{d}z$ is equal to $\int_{0}^{\infty} \text{d}x$ by Cauchy theorem if $R\rightarrow \infty$. Along the path $\Gamma$, $z=R(1+e^{\mathbbm{i}\theta})$  with $\theta\in \left[\pi, 2\pi\right]$. Thus, the real part of $z$ is ensured to be nonnegative, characterizing the energy mode in the environment.  The value of $R$ is responsible for the effective energy modes in the environment, used to simulate the spin dynamics.

Similar to the way in \cite{discretization}, the following transformation is introduced:
\be\label{complextransformation}
d_n&=& \int_{\Gamma}\text{d}z \sqrt{\frac{w(z)}{\mathbbm{i} z}} \eta_n(z)  a_z\nonumber\\
d_n^{\ddag}&=& \int_{\Gamma}\text{d}z \sqrt{\frac{w(z)}{\mathbbm{i} z}} \eta_n(z)  a_z^{\dagger}
\ee
and the inverse
\be\label{bz}
a_z&=&\sqrt{\frac{w(z)}{\mathbbm{i} z}} \sum_{n=0}^{N-1}\eta_n(z) d_n\nonumber\\
a_z^{\dagger}&=&\sqrt{\frac{w(z)}{\mathbbm{i} z}} \sum_{n=0}^{N-1}\eta_n(z) d_n^{\ddag},
\ee
where $a^{\dagger}_z (a_z)$ is the complex generalization of $ a^{\dagger}_x (a_x)$. It is natural to require the invariance of bosonic commutation relation, i.e. $\left[a^{\dagger}_z, a_{z'}\right]=\delta(z -z')$.   $\eta_n(z)$ denotes the polynomial of degree $n$, which can be constructed through  the inner product with respect to weight function $w(z)$,
\be \label{complexinp}
\left\langle f, g \right \rangle_{\Gamma} = \int_{\Gamma} \frac{\text{d}z}{\mathbbm{i} z} w(z) f(z)g(z).
\ee
It is stressed that  the inner product is defined deliberately without complex conjugation to construct the three-term recurrence relation for $\eta_n(z)$ \cite{gautschi}. In this way,  $\eta_n(z)$ can display two features, which are crucial for discretization of Eq. \ref{continuousH}. One is the orthonormality defined as
\be
\left\langle \eta_m, \eta_n \right \rangle_{\Gamma}=\delta_{m, n}.
\ee
The other is the recurrence relation
\be \label{complexrecurrence}
\sqrt{\nu_{n+1}} \eta_{n+1}(z)= \left(z -\mathbbm{i} \mu_n\right) \eta_{n}(z) -\sqrt{\nu_{n}}\eta_{n-1}(z),
\ee
where
\be
\mathbbm{i}\mu_n=\left\langle z \eta_n, \eta_n\right\rangle_{\Gamma},
\nu_n=\frac{A^2_{n-1}}{A_n^2},
\ee
$A_n$ being the coefficient of  $z^n$ in $\eta_n(z)$. With help of these two properties and transformation Eq. \eqref{bz} together, Eq.\eqref{continuousH} can be transformed into the following form
\be\label{Hd}
H&=&\Delta \sigma_+ \sigma_- + \omega_c\sum_{n=1}^{N}\left(\mathbbm{i}\mu_n d_n^{\ddag}d_n + \sqrt{\nu_{n}}d_n^{\ddag}d_{n-1}+ \text{H. c. } \right)+\nonumber\\
&& \sum_{n=1}^{N}\int_{\Gamma}\text{d}z \sqrt{\frac{w(z)}{\mathbbm{i} z}} \eta_n(z)\left[g(z) \sigma_+ d_n + g^*(z) d_n^{\ddagger}\sigma_-\right],
\ee
where $N$ is the highest degree of the polynomial, used in the evaluation. $h(z)$ and $g(z)$ can be obtained directly through replacing $x$ in Eq. \eqref{gh} by $z$.

As a further simplification, let us introduce the new mode operator
\be \label{tildeD}
\widetilde{d}_i&=& \sqrt{w_i} \sum_{n=1}^{N} \eta_n(z_i) d_n,\nonumber \\
\widetilde{d}_i^{\ddagger}&=& \sqrt{w_i} \sum_{n=1}^{N} \eta_n(z_i) d_n^{\ddagger},
\ee
where $z_i$ denotes the root of $\eta_N(z)$, and $w_i$ is the corresponding weight. Both $z_i$ and $w_i$ can be obtained by diagonalizing  the second term in Eq. \eqref{Hd}. For the last term in Eq. \eqref{Hd}, the integration can be approximated by complex Gauss quadrature \cite{gautschi} as
\be
\int_{\Gamma}\text{d}z \sqrt{\frac{w(z)}{\mathbbm{i} z}} \eta_n(z) g(z) &=& \int_{\Gamma}\text{d}z \frac{w(z)}{\mathbbm{i} z} \sqrt{\frac{\mathbbm{i} z}{w(z)}} \eta_n(z) g(z) \nonumber \\
&\simeq&\sum_{i=1}^{N} \sqrt{\frac{\mathbbm{i} z_i}{w(z_i)}} w_i \eta_n(z_i) g(z_i).
\ee
Together with Eqs.\eqref{tildeD}, one finally obtains
\be\label{Hdis}
H_{\text{dis}}&=&\Delta \sigma_+ \sigma_- +\sum_{i=1}^{N} z_i \widetilde{d}_i^{\ddagger}\widetilde{d}_i +\sum_{i=1}^{N} \left(\mathbbm{g}_i \sigma_+ \widetilde{d}_i + \text{h.c.}\right),
\ee
where $\mathbbm{g}_i= \sqrt{\frac{\mathbbm{i} z_i}{w(z_i)}} \sqrt{w_i}g(z_i)$.

Obviously, $H_{\text{dis}}$ is non-Hermitian, and the evolution operator can be written as \cite{brody14}
\be\label{U}
U(t)= \sum_n e^{-\mathbbm{i E}_n t} \ket{n}_R {_L \bra{n}},
\ee
where $\ket{n}_R$ denotes the right eigenfunction of  $H_{\text{eff}}$ with eigenvalue $\mathbbm{E}_n$, and $\ket{n}_L$ denotes the left eigenfunction with eigenvalue $\mathbbm{E}^*_n$. As an exemplification, the survival probability $\left|\inp{\psi(t)}{\psi (0)}\right|^2$ with $\ket{\psi (0)}=\ket{e}\ket{0}^{\otimes N}$ is explored  respectively by exact solution of Eq. \eqref{alphat} and  CDA. The computational error, defined as
\be\label{error}
\text{Error}= \frac{\text{CDA}- \text{Exact}}{\text{CDA}+ \text{Exact}},
\ee
is presented in Fig. \ref{fig:error} for different values of  $s$ and $\eta$ (the explicit plots for the evolution of survival probability are provided by Fig.\ref{fig:exact} in \ref{appendixCDA}). Nonetheless, due to the time-consuming nature of solving Eq. \eqref{alphat},  the exact evaluations are restricted up to $t=10$ with the step length of  $10^{-4}$. For CDA, the weight function chosen is $w(z)=1$, which might cause the appearance of complex eigenmodes in the environment. It should be stressed that choosing $g(z)$ as the weight function will not generate a non-Hermitian effective Hamiltonian. Evidently, the computational errors has a magnitude of $10^{-3}$ in most circumstances, and thus CDA can provides reliable evaluation of the spin dynamics.

\subsection{Introduction of $\tilde{H}_{\text{dis}}$}

It is noted that Fig. \ref{fig:error}(c) shows a significant increase of  error when $s=1$. The explicit calculations reveals a slow convergence to a stable result as $N$ increases. To capture the long-term behavior of spin dynamics, we propose to use
\be\label{tildeH}
\widetilde{H}_{\text{dis}}=\sqrt{H^{\dagger}_{\text{dis}} \cdot H_{\text{dis}}}.
\ee
as an evaluation. In practice,  $H^{\dagger}_{\text{dis}}$ can create a gain effect since its complex eigenvalues have a positive imaginary part. Thus, it is expected to reduce the unwanted decreasing induced by finite $N$. To justify this proposition, we reexamine the spin dynamics of $s=1$ by $\widetilde{H}_{\text{dis}}$.  As shown in Fig. \ref{fig:errorHtilde}(a), $\widetilde{H}_{\text{dis}}$ provides a reliable assessment of $\left|\inp{\psi(t)}{\psi (0)}\right|^2$ for $\eta=0.05$ and $0.1$, with a significant reduction of error by at least a  factor of 10. The explicit illustration for the evolution can be found in Fig. \ref{fig:tildeH} in \ref{appendixCDA}

For $\eta=0.5$. the evaluation of $\tilde{H}_{\text{dis}}$ is reliable only for larger $t$, however it becomes very poor for $t<3$, as shown in Fig. \ref{fig:tildeH}(c).   
Therefor,  we use $H_{\text{dis}}$ for spin dynamics evaluation for times shorter than $\sim 4$ and $\tilde{H}_{\text{dis}}$ for times greater than 4.  It is important to note that $\widetilde{H}_{\text{dis}}$ is  only applicable  when $s=1$, whereas $H_{\text{dis}}$ can provide reliable results for $s=0$ and $0.2$.

To justify the introduction of $\tilde{H}_{\text{dis}}$, we offer the analytical evidence in this section as an addition to the numerical evidence presented in Fig. \ref{fig:convergence} in \ref{appendixCDA}. The objective is to ascertain the long-term behavior of $\alpha(t)$, which can be determined analytically by solving Eq. \eqref{alphap}. It is worth noting that the enduring features of $\alpha(t)$ are determined by the bound state \cite{kofman1994}, which occurs when
\be
\Delta - \int_0^\infty\text{d}\omega \frac{J(\omega)}{\omega} <0.
\ee
For the Ohmic-type environment defined by Eq. \eqref{J},  one gets
\be
\eta>\frac{\Delta}{\omega_c \Gamma(s)}:= \eta_c
\ee
Evidently, $\eta_c=0.1$ when $s=1$ and $\Delta=1, \omega_c=10$. By solving  Eq.\eqref{bs}, a single bound state can be found at $E_b=- 2.369$  when $\eta= 0.5$. At $E_b$, the Laplace transformation $\alpha(p)$ defined in Eq. \eqref{alphap}, can be obtained readily by finding $\lim_{E\rightarrow E_b} \left( E - E_b\right)\alpha(p=\mathbbm{i}E)$. This leads to the approximation of $\alpha(t) \approx -0.739894 e^{- \mathbbm{i} 2.369 t}$ and a resulting survival probability of $\sim 0.547439$. In contrast, the calculation  by $\tilde{H}_{\text{dis}}$ shows a slight oscillation around $\sim 0.54744$ as depicted in Fig. \ref{fig:errorHtilde}(b). This perfect agreement between the analytical and numerical approaches shows unambiguously the ability of  $\tilde{H}_{\text{dis}}$ to predict the steady feature of $\alpha(t)$. 

In the case of no bound state, the integral in Eq. \eqref{bs} becomes divergent, leading to $\alpha(t)$ approaching zero and the survival probability decreasing as $t$ increases. Due to the lack of an analytical expression for $\alpha(t)$, we have to rely on numerical methods. The calculations using $\tilde{H}_{\text{dis}}$ in Fig. \ref{fig:tildeH} (a) and \ref{fig:tildeH}(b) show a reasonably prediction for the steady behavior of survival probability, based on the trend with increasing $N$.    

\begin{figure}[tb]
\center
\includegraphics[width=7cm]{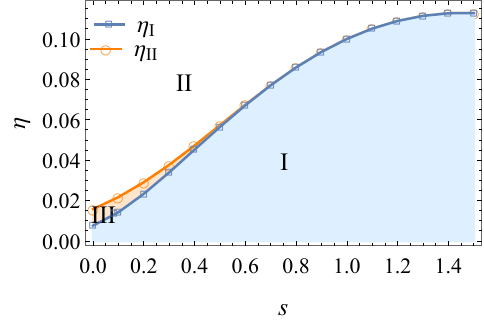}
\caption{(Color online) The phase diagram for $H_{\text{dis}}$ in the single-excitation subspace. All points are determined by exact diagonalization of $H_{\text{dis}}$. Region I:  delocalized phase. Region II:  localized phase. Region III: intermediate phase.  $\eta_I$ denotes the critical value, where the energy gap is zero. $\eta_{II}$ denotes the critical value, where the imaginary part of  ground state function on basis state $\ket{e} \ket{0}^{\otimes k}$ is zero.  For this plot, $R=6, N=1000$ and $\Delta=1, \omega_c =10$ are chosen.  }
\label{fig:phase}
\end{figure}

\begin{figure*}[tb]
\center
\includegraphics[width=16cm]{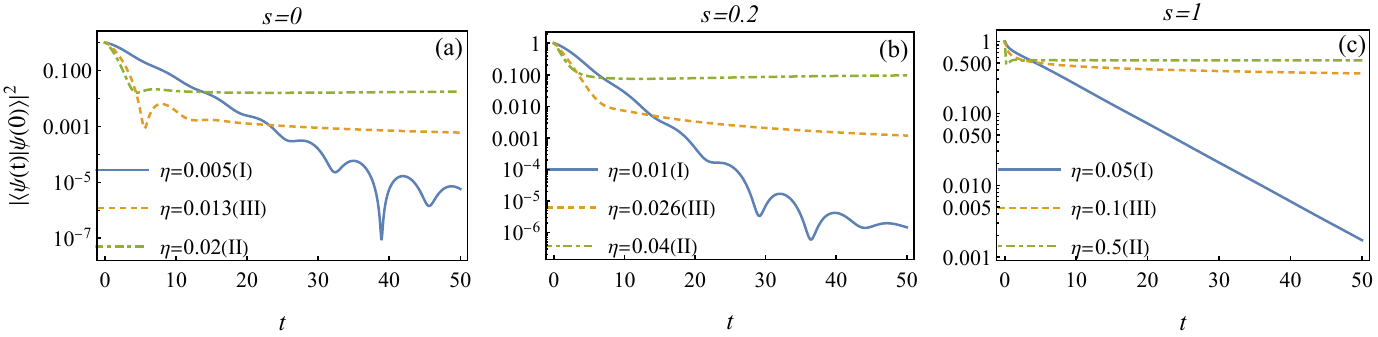}
\caption{(Color online) The evolution of survival probability with time $t$ (in units of $1/\Delta$) is plotted for selected $s$ and $\eta$. For panels (a) and (b), the evolution is obtained by $H_{\text{dis}}$. While for panel (c), $\widetilde{H}_{\text{dis}}$  is used to obtain convergent results for $\eta=0.05, 0.1$. As for $\eta=0.5$, the evolution is determined by the combination of $H_{\text{dis}}$ and $\widetilde{H}_{\text{dis}}$, as stated at the end of Sec. \ref{section:CDA}. The Roman numbers denote the different regions in Fig. \ref{fig:phase}, to which the chosen values of $s$ and $\eta$ belong.  For all plots, $N=2000, R=6, \Delta=1$, and $ \omega_c =10$ are chosen.  }
\label{fig:spindynamics}
\end{figure*}

\section{Quantum Phase Transition in the single-excitation subspace}

Actually, the CDA method transforms the open quantum system into a closed many-body system, allowing for the conventional approach to quantum phase transition \cite{sachdev}  to be applied. An overview to the main result of this section is provided first. After that, the spectrum  of $H_{\text{dis}}$ is studied, with a focus on the energy gap of \ref{fig:tildeH} ground state, which can be used to identify the quantum phase transition. Finally, the ground state function is analyzed in details, which shows a significant change in the population of excitation  with an  increase of the coupling strength $\eta$ and thus indicates an intrinsic transition of spin dynamics.

\subsection{overview to the main result}

The main result is summarized as the phase diagram shown in Fig. \ref{fig:phase}. Evidently, the phase diagram can ge divided into three regions, labeled as I, II and III. Region I denotes the dissipative phase, where the spin dynamics decay rapidly and  excitation is absorbed into the environment. In contrast, region II denotes the localized phase, where excitation may be localized in the spin system by a finite probability. Between these two regions, a third region, named as phase III,  can be found for small $s$. In this region, the spin dynamics shows an intermediate  behavior between the fast decaying and localization. Our calculation shows that the dynamics of the system in this region displays an stretched-like feature. This feature is a result of the competition of the dissipation induced by coupling to the environment and the localization from the opening of energy gap of the ground state. The boundary between region I and II or III is decided by the critical value $\eta_I$ or $\eta_{II}$. At $\eta_I$, the energy gap of ground state opens, while at $\eta_{II}$ the population of excitation of the ground state becomes maximal in the spin degree of freedom. For large $s$,  $\eta_{I}$ and $\eta_{II}$ merge into a single one, at which the spin dynamics shows an intermediate feature. The relevance of $\eta_I$ and $\eta_{II}$ to $N$ and $R$ is studied in \ref{appendixQPT}

The survival probability $\left| \inp{\psi(t)} {\psi(0)}\right|^2$ with $\ket{\psi (0)}=\ket{1} \ket{0}^{\otimes k}$ is evaluated to  illustrate  the distinct dynamics in three regions. As shown in Fig. \ref{fig:spindynamics}, dependent on the values of $s$ and $\eta$ belonging to different regions in  Fig. \ref{fig:phase}, $\left| \inp{\psi(t)} {\psi(0)}\right|^2$ can display fast decaying and localization respectively.  To display the intermediate  dynamics in region III, a numerical fitting for the long-term evolution of $\left| \inp{\psi(t)} {\psi(0)}\right|^2$ is provided in Fig. \ref{fig:stretched} in \ref{appendixQPT}, based on the fitting function $B\exp\left(-At^{\beta}\right)$. It is found that the exponential index $\beta$ has the magnitude of order $10^{-1}$, which clearly demonstrates the stretched-like feature of the evolution.

\subsection{The spectrum of $H_{\text{dis}}$}

\begin{figure*}[tb]
\center
\includegraphics[width=16cm]{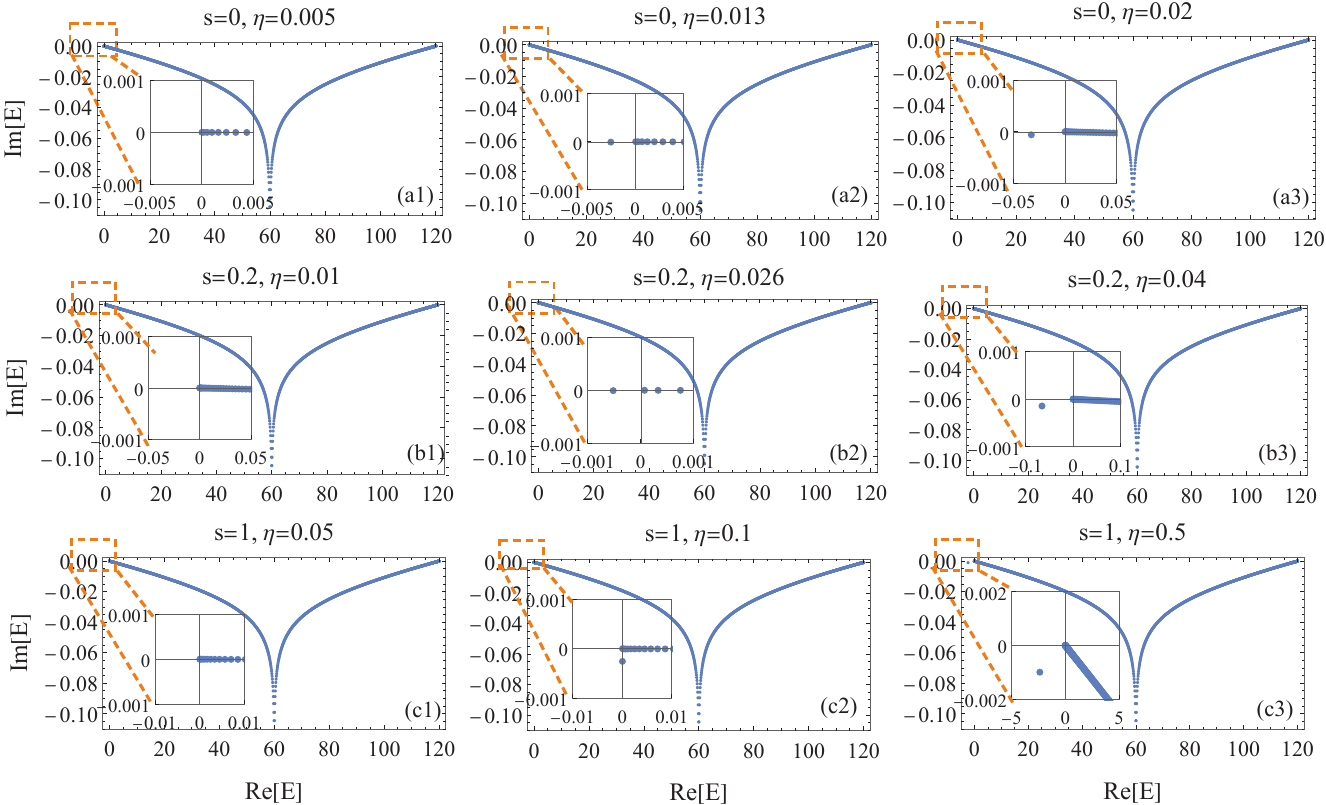}
\caption{(Color online)Plots of the spectrum (in units of $\Delta$) of $H_{\text{dis}}$ for different $s$ and  $\eta$. For all plots, $N=2000, R=6, \Delta=1$, and $ \omega_c =10$ are chosen. }
\label{fig:spectrum}
\end{figure*}

To understand the phase diagram, the spectrum of $H_{\text{dis}}$ is studied in this subsection, obtained by exactly  diagonalizing $H_{\text{dis}}$. As shown in Fig. \ref{fig:spectrum}, the eigenvalues exhibit a non-positive imaginary part, characterizing the dissipative spin dynamics. Furthermore, their  real parts are confined at interval $\left(0, 120\right)$. This is a consequence of the CDA approach, where the continuum of the environment may be discretized as a finite spectrum bound at interval $\left(0, 2R \omega_c\right)$. 

As for the non-Hermitianity of $H_{\text{dis}}$, the ground state can be defined as the right eigenfunction with the minimal real part of its eigenvalue. For small $\eta$, the ground state would correspond to the energy level with the real part close to zero, as shown by the left column in Fig.\ref{fig:spectrum}.  However when $\eta$ increases beyond a critical value $\eta_I$ related to $s$, a single level emerges  in the region $\text{Re}[E]<0$, as shown by the middle and right columns in Fig.\ref{fig:spectrum}. In this case, the ground state becomes  separated from the level band by a energy gap, which is equal to the real part of its eigenvalue. In addition, the gap increases  with increment of $\eta$.

Physically, the single ground state corresponds to the photon-bound state, which has been known to be associated with the localized dynamics of a system \cite{kofman1994,bs}. Nevertheless, as shown in Fig. \ref{fig:spindynamics}, the opening of the energy gap does not guarantee the emergence of localization in the dynamics,  a stretched-like decaying observed when $\eta>\eta_{I}$. Additionally,   localized dynamics can only be observed when $\eta$ surpasses a second critical value. Consequently,  to understand the intricate behaviors, it is necessary to inspect the ground state function.

\begin{figure*}[tb]
\center
\includegraphics[width=16cm]{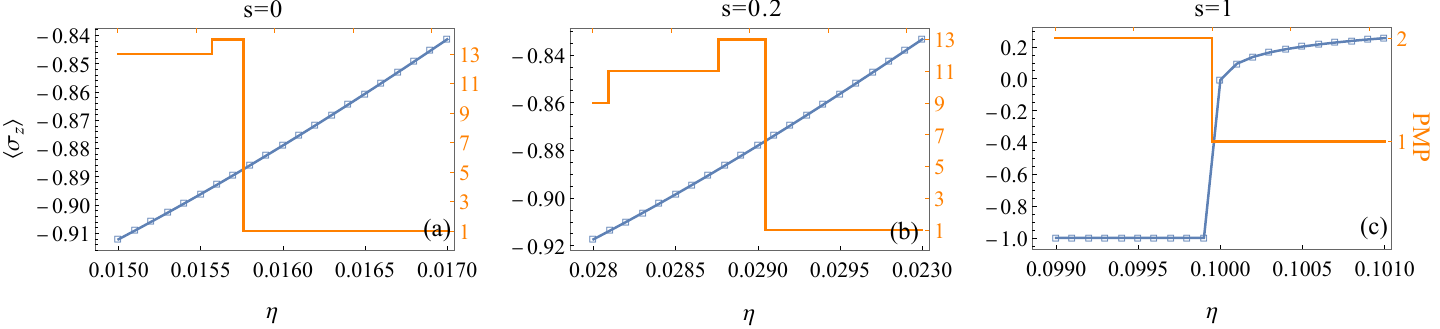}
\caption{(Color online) The PMP(orange line) and  average value of $\sigma_z$ (blue line) for the ground state of $H_{\text{dis}}$ vs. $\eta$. For P.M.P., the arabic numbers denote the basis state $\ket{k}$. The parameters are chosen the same in Fig. \ref{fig:spectrum} }
\label{fig:sigmaz}
\end{figure*}

\subsection{The ground state function}

The ground state function can be obtained directly by solving $H_{\text{dis}}$, which can be written  formally as  $\ket{g}=\alpha \ket{e} \ket{0} + \ket{g} \left(\sum_{k} \beta_k(t) \ket{1}_k\right)$. Here, $\ket{e}\ket{0} \equiv \ket{1}$ denotes excitation  in the spin, while $\ket{g}\ket{1}_k \equiv \ket{k} (k=2, 3, \cdots, N+1)$ denotes the excitation in the $k$-th mode of the environment. The population of excitation on $\ket{1}$ has a significant influence on the localized evolution of $\left| \inp{\psi(t)} {\psi(0)}\right|^2$. In Fig. \ref{fig:sigmaz}, the position of  maximal population (PMP) on the basis state $\ket{k}$ is explored for the ground state when $s=0, 0.2$ and $1$. For the three situations,  the energy gap of ground state opens at critical point $\eta_I=0.0076, 0.02335 $ and $0.1$ respectively, as illustrated in Fig.\ref{fig:phase}. However, as shown in Fig. \ref{fig:sigmaz} (a) and (b),  PMP at $\ket{1}$ can happen only when $\eta$ is beyond a critical value, labeled as $\eta_{II}$. This can explain the evolution illustrated in Fig. \ref{fig:spindynamics}(a) and \ref{fig:spindynamics}(b), in which the localized behavior of $\left| \inp{\psi(t)} {\psi(0)}\right|^2$ can occur only for $\eta$  considerably greater than $\eta_I$. Thus, the stretched-like dynamics is a result of  the competition between the energy gap, which  protects excitation against the spontaneous emission into the environment, and the intrinsic dissipation in the ground state function, which does not support PMP at $\ket{1}$.

For $s=1$,  $\eta_{II}$ is found to be almost equal to $\eta_I$.  Thus, the stretched-like evolution can be found at critical point $\eta_I=0.1$. Additionally, Fig. \ref{fig:RandN} (c) in \ref{appendixQPT} reveals that $\eta_{II}$ is smaller than $\eta_I$. This difference, however, tends to diminish when $N\rightarrow \infty$. A similar picture can also be found for other values of $s$.

In addition, the average value $\langle \sigma_z \rangle$ for the ground state  is evaluated to detect phase transition. As shown by Fig. \ref{fig:spindynamics}, $\langle \sigma_z \rangle$ remains continuous for $s=0, 0.2$, yet a jump is observed at $\eta=0.1$ for $s=1$, implying that the transition for $s=1$ may be distinct from $s=0, 0.1$.

Conclusively, we  demonstrated  the presence of a second critical point $\eta_{II}$ by analyzing the ground state function. For small $s$, $\eta_{II}$ is distinct from $\eta_I$, and  an intermediate region can be identified  where  the spin dynamics shows stretched-like behavior.  As $s$ increases,  $\eta_{II}$ tends to be equal to $\eta_I$, and the  stretched-like spin dynamics can happen at a critical point. Furthermore, the two scenarios can be distinguished by evaluating $\langle \sigma_z \rangle$ for the ground state. If $\eta_{II}$ does not equal to $\eta_I$,  $\langle \sigma_z \rangle$ is continuous at $\eta_{II}$. Conversely, if $\eta_{II}$ tends to be equal to $\eta_I$, $\langle \sigma_z \rangle$ exhibits a jump c at $\eta_{II}$.

\begin{figure*}[tb]
\center
\includegraphics[width=16cm]{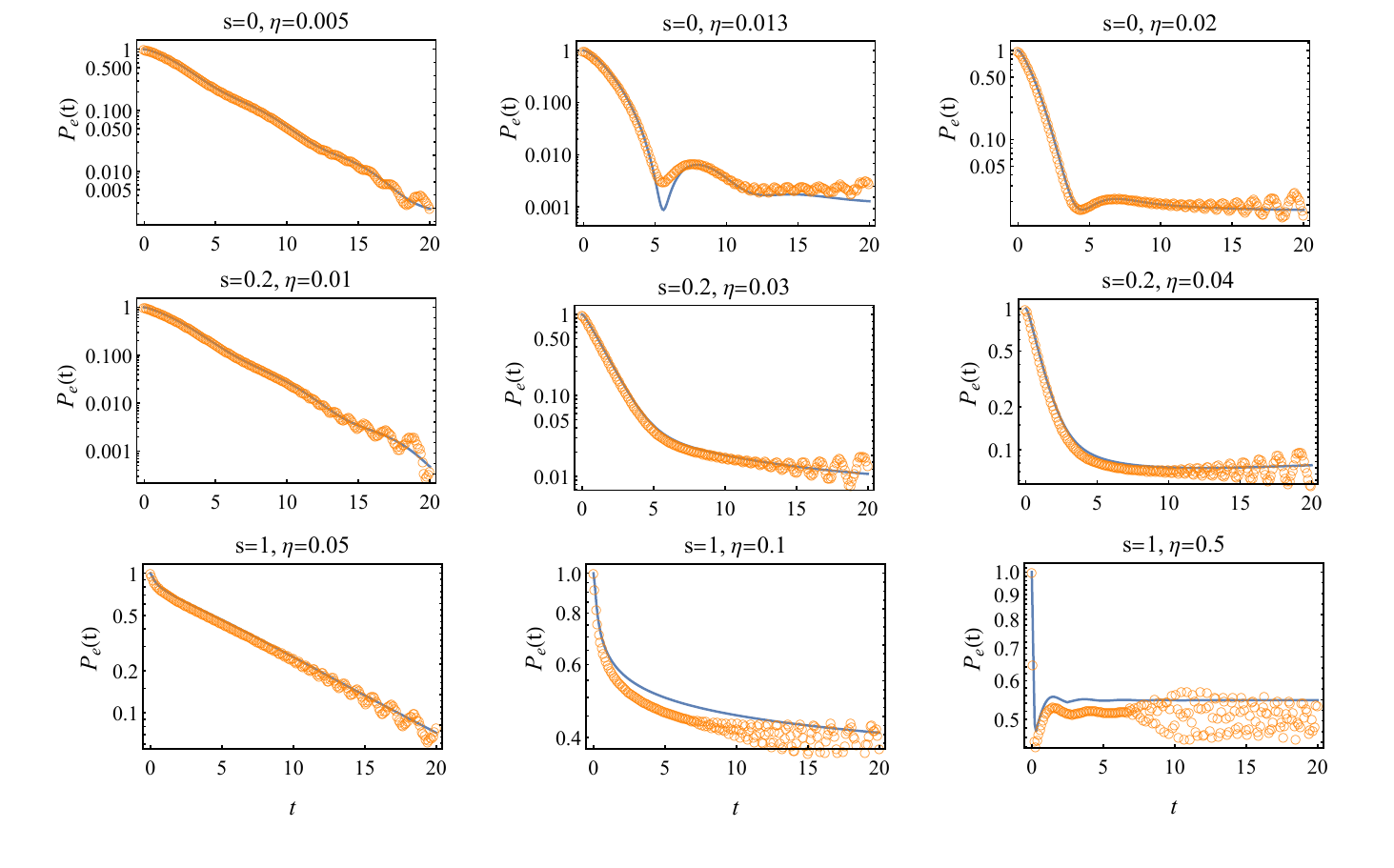}
\caption{(Color online) The comparative plots of spin dynamics in the double-excitation subspace (empty circle) and single-excitation subspace (solid line). The spin dynamics in single-excitation subspace is evaluate by $H_{\text{dis}}$ for $s=0, 0.2$ and $\widetilde{H}_{\text{dis}}$ for $s=1$. For the calculation in  double-excitation subspace, $\widetilde{H}_{\text{dis}}$ is solved by choosing  $N=200$  to obtain the convergent result.  For all plots,  $R=6, \Delta=1, \omega_c =10$ are chosen. The evolution time $t$ is in units of $1/\Delta$. }
\label{fig:double}
\end{figure*}

\section{spin dynamics in the double-excitation subspace}
In the case of multiple excitations, the spin dynamics is difficult  to study due to the absence of a compact equation similar to Eq. \eqref{alphat}. CDA, however,  provides an efficient approach  to  this issue. In principle, the dynamics can be obtained  by exactly diagonalizing $H_{\text{dis}}$ in the multiple excitation space. However, the dimension of $H_{\text{dis}}$ increase drastically  as the number of excitations, making the exact treatment  highly demanding. Therefore, this section focuses only on the spin dynamics in the  double-excitation subspace.

The wave function in the double-excitation subspace can be expressed as
\be
\ket{\phi(t)}=\ket{e} \sum_{k=1}^{N}\alpha_k (t) a^{\dagger}_{k} \ket{0}^{\otimes N} + \ket{g}\sum_{k\leq k'}\beta_{k, k'}(t) a^{\dagger}_{k} a^{\dagger}_{k'}\ket{0}^{\otimes N}
\ee
Evidently,  $\ket{e}$ is correlated to the status of the environment. To characterize the spin dynamics,
\be\label{pe}
P_e(t)= \sum_{k=1}^{N} \left| \alpha_k (t) \right|^2.
\ee
is evaluated for arbitrary time $t$, which denotes the probability of the single excitation saved  by the spin. The initial state is chosen as $\ket{\phi (0)}=1/\sqrt{N} \ket{e} \sum_{k=1}^{N} a^{\dagger}_{k} \ket{0}^{\otimes N} $ to avoid the dependence on the special mode in the environment. $N=200$ is chosen for the simulation, which corresponds to the Hilbert space of dimension $\sim 2.0 \times 10^5$.  For larger $N$, the calculation becomes too consumptive to implement. To offset the recurrence induced by finite $N$, $\widetilde{H}_{\text{dis}}$ is used to find convergent $P_e(t)$ in this evaluation, which is illustrated by the empty circles in Fig. \ref{fig:double}. Because of finite $N$, the calculation becomes fluctuated when $t>\sim 10$. For $s=1$, the calculation shows a slow convergency, which induces strong fluctuation and large computational error.

It is observed  that $P_e(t)$ displays a similar evolution as  $\left|\inp{\psi(t)}{\psi (0)}\right|^2$ in the single excitation subspace. This has been verified  further by the Gauss quadratures method in real space \cite{discretization}, as illustrated by Fig. \ref{fig:realdouble} in \ref{appendixrealGQ}. This observation would be attributed to the fact  that the discretized environment corresponds to an interaction-free bosonic system, in which the excitations can be populated   freely at the mode of the environment  without any interference. Thus, the decay of excitation in spin would be independent of the status of the environment. This  implies that the phase diagram in single-excitation subspace would be robust against the increase of energy in the total system.

We try to find an explanation of spin dynamics by analyzing the spectrum of $H_{\text{dis}}$. However, the spectrum of $H_{\text{dis}}$ does not provide sufficient evidence to explain the spin dynamics in terms of the ground state, as is possible in the single-excitation subspace. Calculations show that the energy gap can remain finite even when $\eta<\eta_I$, and the ground state can have positive energy in such cases. Furthermore, the number of eigenvalues with negative real parts depends on the value of $\eta$. Consequently, the bound state is absent in the multiple-excitation subspace, and the dynamics of spin cannot be explained from the single-excitation perspective.

\section{Conclusion}

In conclusion, the quantum phases in the extended Jaynes-Cummings model was investigated in this paper. To simulate the open dynamics of spin, a non-Hermitian effective Hamiltonian $H_{\text{dis}}$ was  constructed using complex discretization approximation for the environment. The validity of this approach was confirmed  by examining the spin dynamics in the single-excitation subspace, which was determined exactly by solving Eq. \ref{alphat}. By exactly diagonalizing $H_{\text{dis}}$, the spin dynamics could be further examined by analyzing the ground state, which revealed an eigenvalue with the smallest real part. It was found that the spin dynamics in single-excitation subspace was strongly related to two key properties of the ground state,  the energy gap, defined as the negative real part of its eigenvalue, and the PMP of ground state on the basis state. When the energy gap is zero, the spin dynamics, measured by the survival probability of initial state $\ket{\psi (0)}=\ket{1} \ket{0}^{\otimes k}$, decays rapidly. Conversely, if the energy gap is finite, two distinct spin dynamics can be observed, determined by PMP. If  PMP occurs at $\ket{1}$,  the survival probability stabilizes eventually at a finite value, which indicates the localization of excitation in the spin. If this does not occur, a stretched-like dynamics was observed. This picture is a result of the competition between the energy gap, which  protects excitation against the spontaneous emission into the environment, and the intrinsic dissipation in the ground state function, which does not support PMP at $\ket{1}$.

By detecting the occurrence of an energy gap and PMP at basis state $\ket{1}$, the critical point $\eta_I$ or $\eta_{II}$ can be identified, which separates the localized phase from  the dissipative phase  or  the intermediated phase respectively. Additionally, both $\eta_i$ and $\eta_{II}$ display relevance to the value of $s$. For $0< s<\sim 0.6$, $\eta_{II}$ is greater than $\eta_I$, which leads to the identification of the intermediated phase. While for $s>\sim 0.6$, $\eta_{II}$ tends to be equal to  $\eta_I$.

The spin dynamics was also investigated  in the double-excitation subspace by evaluating $P_e(t)$ defined in Eq. \eqref{pe}. Interestingly,  the same evolution as the survival probability in single-excitation subspace is observed. This feature would be attributed to the fact that the discretized environment corresponds to the interaction-free boson model, which enables the excitations  to  travel in the environment without the interference. Importantly, it is found that the bound state cannot be constructed in this case, and the dynamics of spin cannot be explained from the single-excitation perspective.


\section*{ACKNOWLEDGEMENTS}
H.T.C. acknowledges the support of the Natural Science Foundation of Shandong Province under Grant No. ZR2021MA036. Y. A. Y. acknowledges the support of the National Natural Science Foundation of China (NSFC) under Grant No. 21973036.  M.Q. acknowledges the support of NSFC under Grant No. 11805092 and the Natural Science Foundation of Shandong Province under Grant No. ZR2018PA012. X.X.Y. acknowledges the support of NSFC under Grant No. 12175033 and the National Key R$\&$D Program of China (Grant No. 2021YFE0193500).

\renewcommand\thefigure{A\arabic{figure}}
\renewcommand\theequation{A\arabic{equation}}
\renewcommand\thesection{Appendix \Roman{section}}
\setcounter{equation}{0}
\setcounter{figure}{0}
\setcounter{section}{0}

\section{Supplementary materials on  CDA approach to spin dynamics in Eq. \ref{H} }\label{appendixCDA}

\begin{figure*}[tb]
\center
\includegraphics[width=16cm]{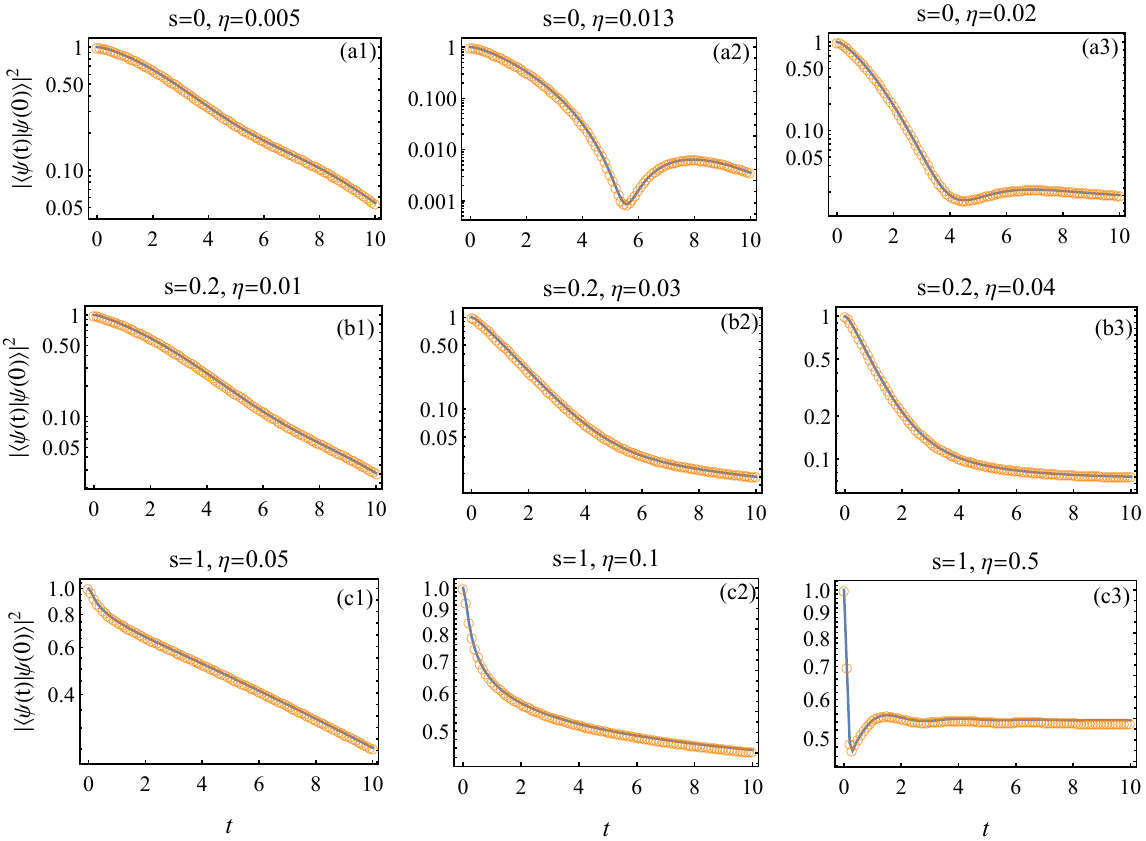}
\caption{(Color online) The comparative plots for the survival probability $\left|\inp{\psi(t)}{\psi (0)}\right|^2$  with $\ket{\psi (0)} = \ket{e} \ket{0}^{\otimes k}$ for different $s$ and  $\eta$. The solid line depict the exact result, whereas the empty circles represent results obtained by CDA.   For all plots, $N=2000, R=6, \Delta=1$, and $ \omega_c =10$ are chosen. The evolution time $t$ is in units of $1/\Delta$. }
\label{fig:exact}
\end{figure*}

\begin{figure*}[tb]
\center
\includegraphics[width=15.5cm]{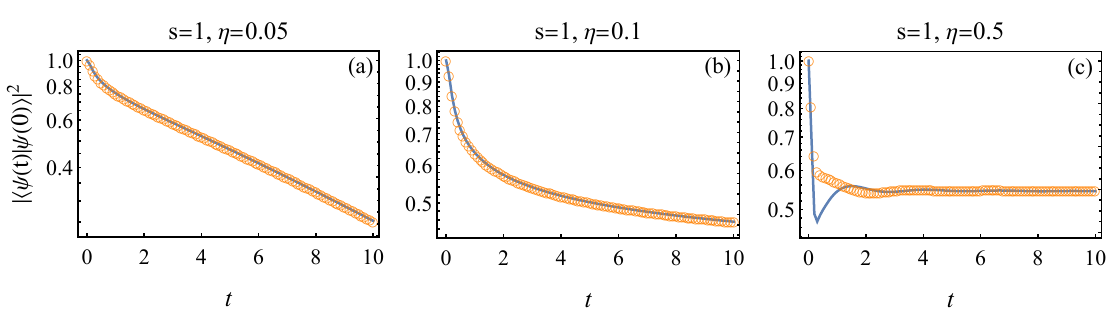}
\caption{(Color online) The comparative plot between the exact results (solid line) and the numerics (empty circle) obtained by $\widetilde{H}_{\text{dis}}$.  The other parameters are same to those in Fig. \ref{fig:exact}. The evolution time $t$ is in units of $1/\Delta$. }
\label{fig:tildeH}
\end{figure*}

\begin{figure*}[tb]
\center
\includegraphics[width=16cm]{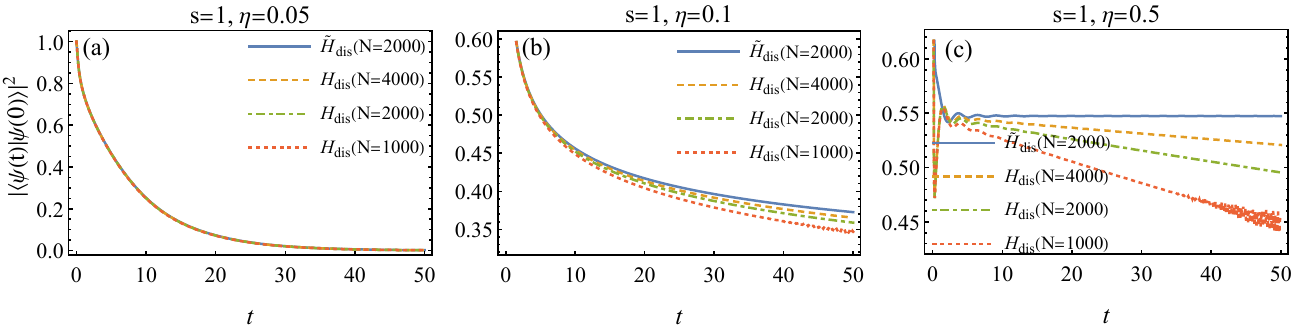}
\caption{(Color online) Plots for CDA approach to the long-term feature of $\left|\inp{\psi(t)}{\psi (0)}\right|^2$   when $\widetilde{H}_{\text{dis}}$ and $H_{\text{dis}}$ are used respectively. It is evident that the numerics by $H_{\text{dis}}$ would become convergent to the results of $\widetilde{H}_{\text{dis}}$  when $N$ increases. The evolution time $t$ is in units of $1/\Delta$.}
\label{fig:convergence}
\end{figure*}

Figure \ref{fig:exact} illustrates explicitly the evolution of the survival probability $\left|\inp{\psi(t)}{\psi (0)}\right|^2$ for different values of $s$ and $\eta$. The solid line is decided by solving Eq. \eqref{psit} with step length $\delta t=10^{-4}$, and the empty circles represent results obtained by CDA.  A slight difference can be observed for $s=1, \eta=0.5$.  Figure \ref{fig:tildeH} illustrates the results obtain by $\widetilde{H}_{\text{dis}}$, which shows obvious deviation from the exact results at short term evolution when $\eta=0.5$.

To justify $\widetilde{H}_{\text{dis}}$,  the long-term feature of $\left|\inp{\psi(t)}{\psi (0)}\right|^2$ is evaluated respectively by $\widetilde{H}_{\text{dis}}$ and $H_{\text{dis}}$, as shown in Fig. \ref{fig:convergence}. Although the convergency of the calculation is rapid for $\eta=0.05$,  the evaluation for $s=0.1, 0.5$ converges slowly to a stable result with the increase of $N$. By the observation of convergence, it is reasonable to consider the results by $\widetilde{H}_{\text{dis}}$ as stable. This conjecture can further be verified by Fig. \ref{fig:convergence} (a), in which both $\widetilde{H}_{\text{dis}}$ and $H_{\text{dis}}$ give  the same predication of spin dymamics.


\section{Supplementary materials  about the quantum phase transition in  single-excitation subspace} \label{appendixQPT}

\begin{figure*}[tb]
\center
\includegraphics[width=16cm]{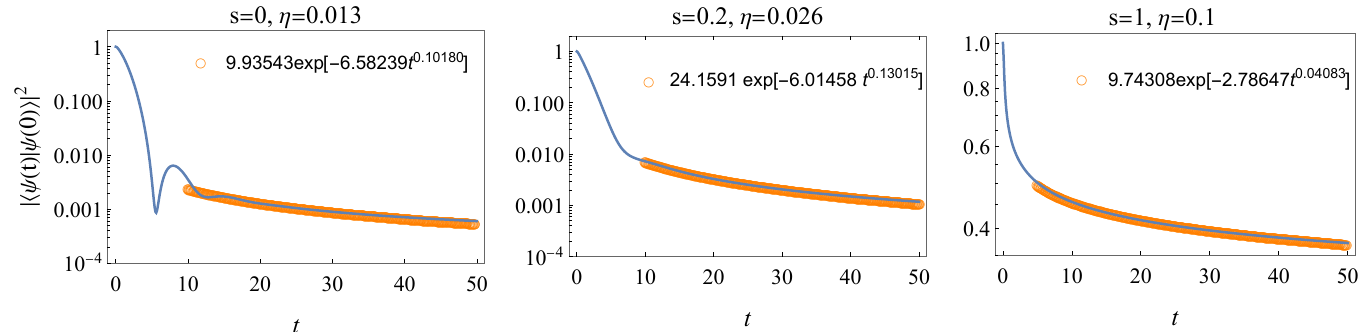}
\caption{(Color online) The numerical fitting for the stretched-like evolution, shown in Figs. \ref{fig:spindynamics}. The evolution time $t$ is in units of $1/\Delta$.}
\label{fig:stretched}
\end{figure*}

\begin{figure*}[tb]
\center
\includegraphics[width=16cm]{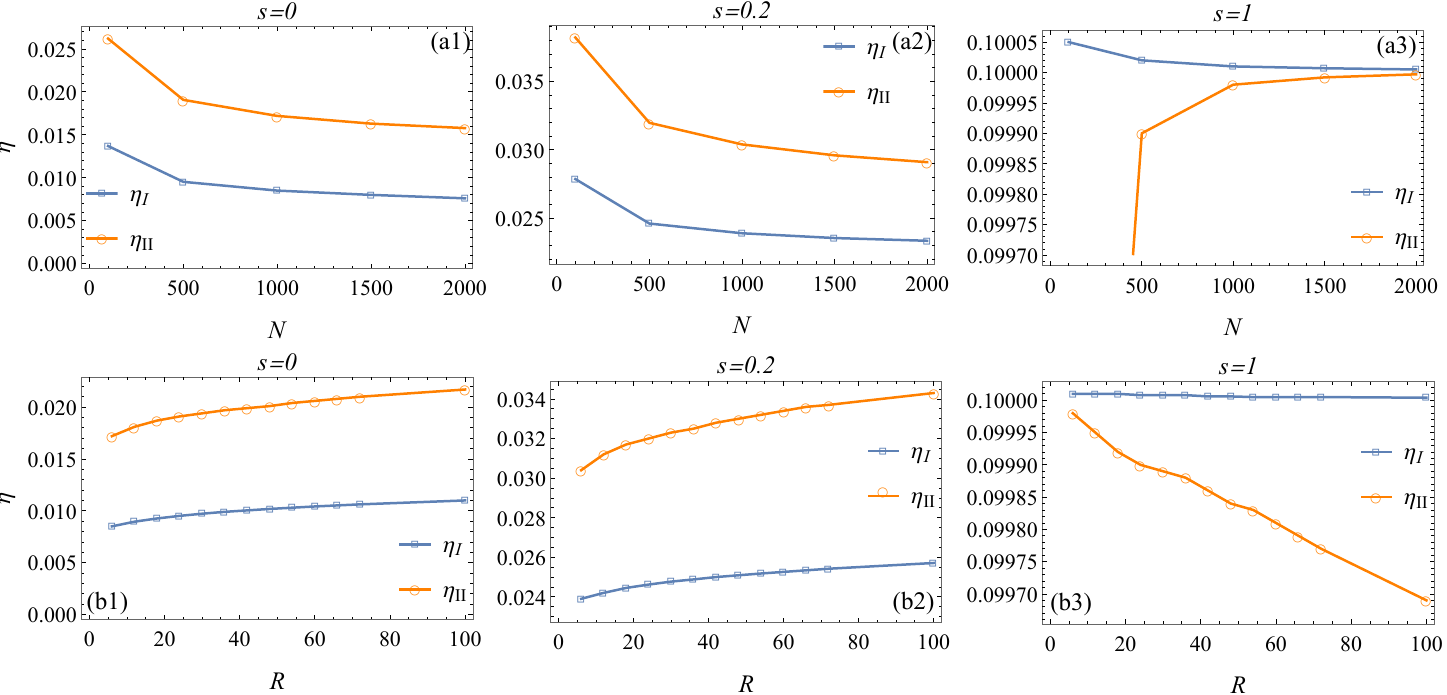}
\caption{(Color online) $\eta_I$ and $\eta_{II}$ vs. the parameter $R$ and $N$. For panels (a1)-(a3), $R=6$ is chosen. While for panels (b1)-(b3), $N=1000$ is chosen. For all plots, $\Delta=1, \omega_c =10$ are chosen.   }
\label{fig:RandN}
\end{figure*}

Figure \ref{fig:stretched} illustrates the fitting  for the stretched evolution of  survival probability, observed in Fig. \ref{fig:spindynamics} for $\eta=0.013, 0.026, 0.1$. The fitting function is chosen as $B\exp\left(- A t^{\beta}\right)$, in which $A, B$ and $\beta$ are real and non-negative. Evidently, $\beta$ has a magnitude of order $10^{-1}$, and thus demonstrates clearly the stretched-like behavior.

In Fig. \ref{fig:RandN}, the relevance of $\eta_I$ and $\eta_{II}$ to the computational parameters $R$ and $N$ is presented. For $N$, it is observed that the values of both $\eta_I$ and $\eta_{II}$ reach a steady state with the rise of $N$.  However,  for $s=1$,  $\eta_I$ and $\eta_{II}$ are observed  to merge into a single value close to $0.1$. On the other hand, with the increase of $R$, $\eta_I$ and $\eta_{II}$ increase very slowly for $s = 0$ and $s = 0.2$. For $s = 1$, $\eta_{II}$ decreases gradually, while $\eta_I$ stays constant. These features are a result of the fact that an increase of $R$ can improve the accuracy of computation, but is not favorable for the enhancement of   efficiency.

\section{Approaching spin dynamics in double-excitation subspace by Gauss quadrature in real space}\label{appendixrealGQ}

\begin{figure*}[tb]
\center
\includegraphics[width=16cm]{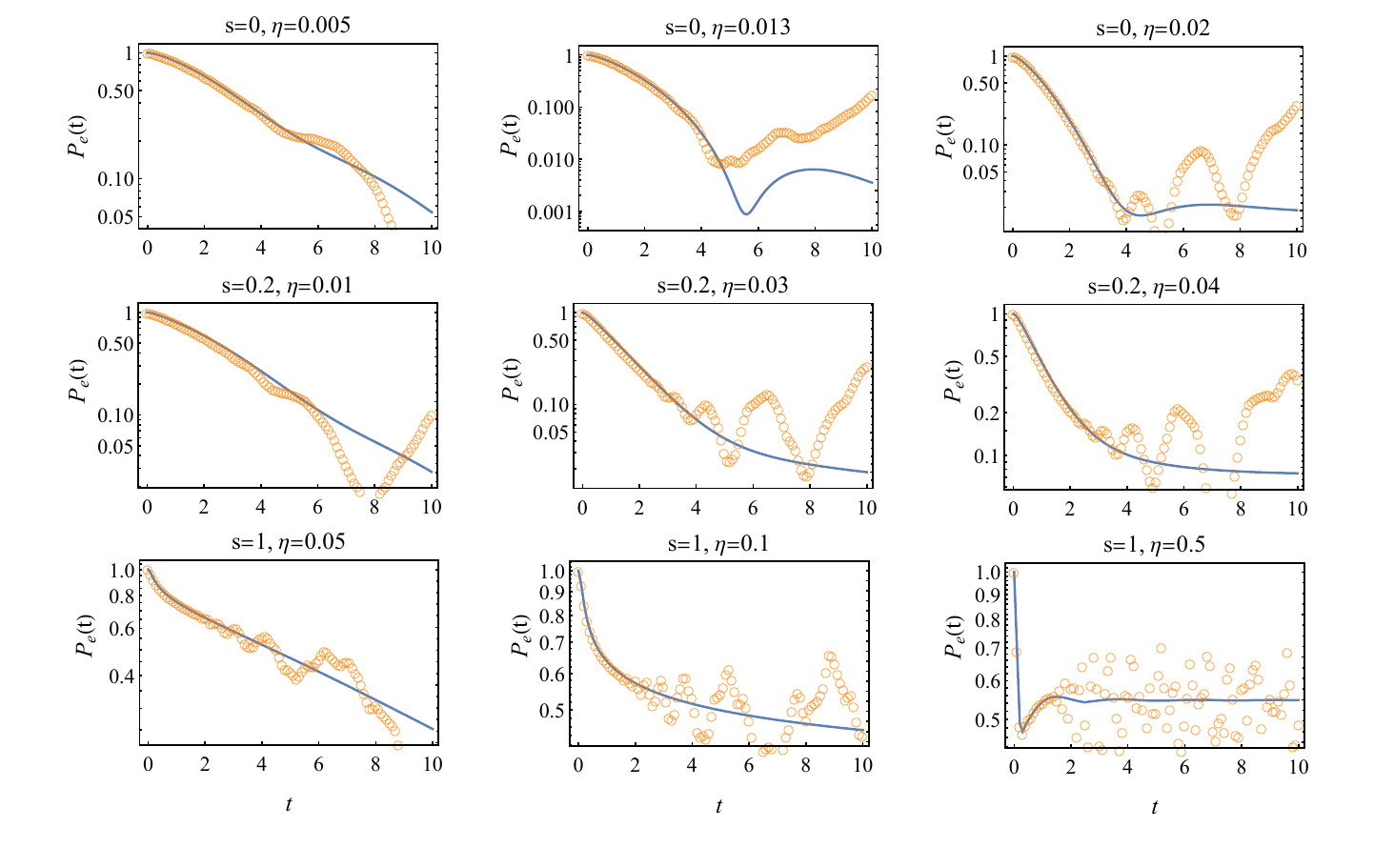}
\caption{(Color online) The reevaluation  of spin dynamics in the double-excitation subspace (empty circle) by Gauss quadratures method.  The solid line corresponds to the result of $\left|\inp{\psi(t)}{\psi (0)}\right|^2$ in single-excitation subspace. For all plots, $\Delta=1$ and $\omega_c=10$ are chosen.}
\label{fig:realdouble}
\end{figure*}

In Fig. \ref{fig:realdouble}, $P_e(t)$ is evaluated using the Gauss quadrature method \cite{discretization}. The weight function $w(x)= \left(x/\omega_c\right)^{s}e^{-x/\omega_c}$ is chosen to construct the chain Hamiltonian and the number of lattice sites is set to $N=200$. It is evident that $P_e(t)$ has a similar evolution as the survival probability $\left|\inp{\psi(t)}{\psi (0)}\right|^2$ in single-excitation subspace. However, in contrast to the CDA approach,  strong fluctuations due to finite $N$ appear when $t >\sim 2$, which limits the application of Gauss quadrature in the simulation of long-term dynamics in open quantum systems.

\end{document}